\documentclass[12pt,a4paper,final]{iopart}
\usepackage{iopams}  
\usepackage{graphicx}
\usepackage{dcolumn}  
\usepackage{bm}       
\usepackage{amssymb} 
\usepackage{braket}
\usepackage{mathcomp}
\usepackage{cite}
\usepackage{braket}

\usepackage{bbold}
\expandafter\let\csname equation*\endcsname\relax
\expandafter\let\csname endequation*\endcsname\relax 
\usepackage{amsmath}
\usepackage{physics}

\begin{document}

\title[Preparing an article for IOP journals in  \LaTeXe]{ Unitary time-reversal on non-orientable spacetimes }
\vspace{3pc}

\author{Ovidiu Racorean}
\address{General Direction of Information Technology}
\address{Banul Antonache str. 52-60, Bucharest, Romania}
\ead{ovidiu.racorean@mfinante.gov.ro}
\vspace{3pc}

\begin{abstract}
\vspace{1pc}

Time reversal symmetry occupies a distinctive role in quantum mechanics, fundamentally requiring an anti-unitary operator to ensure a physically consistent representation. As such, the time reversal operator combines a unitary transformation with complex conjugation, enabling the necessary inversion of the imaginary unit that appears in quantum commutation relations and dynamical equations. Attempts to represent time reversal as a purely unitary operation encounter fundamental contradictions, including violations of canonical commutation relations and issues with the positivity of energy spectra. However, recent advances in quantum gravity and black hole physics reveal that in spacetimes with non-orientable topology—where a global temporal orientation is not well defined—time reversal may be realized by a purely unitary operator. Such non-orientable geometries connect two asymptotically spacetimes with opposite time directions, thereby encoding time reversal topologically and removing the need for complex conjugation. In this work, we explore the deep connection between spacetime orientability and the nature of the time reversal operator, demonstrating that orientable spacetimes require anti-unitary time reversal consistent with conventional quantum theory, while non-orientable spacetimes allow unitary time reversal operators consistent with negative energy states. 

\end{abstract}



\maketitle

\section{Introduction}
Time reversal symmetry ($T$-symmetry) holds a unique and subtle position in quantum mechanics compared to other symmetries. Unlike spatial symmetries, which are represented by unitary operators acting linearly on the Hilbert space, time reversal fundamentally requires an anti-unitary operator \cite{Roberts1, Roberts2} to capture its complete and physically consistent representation. This anti-unitary operator combines a unitary transformation with complex conjugation, ensuring the correct inversion of the imaginary unit $ i $ that appears in canonical commutation relations and quantum dynamical equations. Such an anti-unitary formulation preserves crucial physical properties, including probability conservation and the proper transformation of momentum, spin, and energy under time inversion.

Naively attempting to represent the time reversal as a purely unitary operation leads to fundamental contradictions in quantum mechanics. For example, a unitary operator leaves the imaginary unit $ i $ unchanged, which violates the canonical commutation relations \cite{Hatano, Lombardi} because time reversal must invert momentum while leaving position invariant. Furthermore, if the time reversal were unitary, it would imply an unphysical inversion of energy spectra \cite{Henry1, Henry2, Petit1, Petit2} that are bounded below, violating stability conditions. Consequently, the anti-unitary time reversal operator emerges as a necessary structure in quantum theory, aligning with Wigner’s theorem \cite{Wigner} that mandates symmetry operations be either unitary or anti-unitary.

Nevertheless, recent developments in quantum gravity and black hole physics highlight scenarios where the traditional anti-unitary description may be generalized. In particular, spacetimes with non-orientable topology—where a global temporal orientation cannot be consistently defined—allow for an alternative realization of time reversal through a purely unitary operator. Such non-orientable spacetimes arise in some particular wormhole classes that connect regions with reversed temporal flows such as wormholes in scalar–tensor gravity \cite{Lobo}, $PT$-symmetric wormholes \cite{Zejli} or non-orientable BTZ black holes \cite{Racorean}.  This non-orientability encodes \cite{Bielinska, Hadley} an inherent reversal of temporal orientation topologically, removing the need for complex conjugation in the time reversal operator.

This paper explores the profound interplay between spacetime topology, specifically orientability, and the nature of the time reversal operator in quantum mechanics. We analyze the mathematical and physical distinctions whereby orientable spacetimes require anti-unitary time reversal symmetry, while non-orientable spacetimes permit unitary time reversal operators. Therefore, the nature of the time reversal operator—whether anti-unitary or unitary—is deeply intertwined with the global topological features of spacetime. This paper explores this interplay in depth. It reveals that the non-relativistic Schrödinger equation, defined on orientable spacetimes, requires anti-unitary time reversal, whereas the relativistic Dirac equation, applicable to non-orientable spacetimes, supports a unitary time reversal operator compatible with negative energies. 

These insights open new avenues for understanding time reversal symmetry, negative energy states, and their connection to spacetime topology in quantum gravity and high-energy physics.

\section{Unitary and anti-unitary time reversal}

Time reversal symmetry ($T$-symmetry) in quantum mechanics is unconventional compared to other symmetries because it fundamentally requires an anti-unitary operator for its complete and physically consistent representation. The concept of time reversal symmetry in quantum mechanics traditionally involves an anti-unitary operator, which combines a unitary transformation with complex conjugation to invert the direction of time evolution while preserving canonical commutation relations and physical observables. However, intriguing theoretical investigations have explored the possibility of representing time reversal by a purely unitary operator \cite{Petit2, Callender}, leading to novel implications. 

Unitary operators commonly represent symmetry transformations that preserve probabilities and phases without complex conjugation. For example, spatial parity ($P$) is represented by a unitary operator that reverses position and momentum but leaves the imaginary unit $ i $ unchanged:

\begin{equation}
PxP^{-1}=-x, PpP^{-1}=-p, PiP^{-1}=i.
\end{equation}

Attempting to represent time reversal as a purely unitary operation creates immediate physical problems. Time reversal is expected to reverse momentum but leave position unchanged:

\begin{equation}
TxT^{-1}=x, TpT^{-1}=-p.
\end{equation}

If $T$ were unitary, it would leave i intact:

\begin{equation}
TiT^{-1}=i.
\end{equation}

This leads to a contradiction because the canonical commutation relation 
\begin{equation}
[x,p]=i\hslash,
\end{equation}

would not be preserved under this transformation: 

\begin{equation}
[x, -p]=-i\hslash \neq i\hslash,
\end{equation}

violating a core algebraic structure of quantum mechanics.

Moreover, if time reversal were unitary, it would reverse the sign of energy (the time component of four-momentum) just as spatial parity reverses the momentum sign. This is physically untenable since energy spectra are bounded below (positive definite). Because energy eigenstates involve phases of the form $e^{-iEt}$, the direction of time reversal must also include reversing the sign of $i$.

Consequently, the time reversal operator must include complex conjugation which sends $i \to -i $, making it anti-linear and anti-unitary:

\begin{equation}
T=UK,
\end{equation}

where $U$ is unitary and $K$ is complex conjugation.

However, from a formal standpoint, one can conceive a "unitary time reversal" as a mathematical object that reverses time coordinate parametrization classically or naïvely, applying a linear unitary operation without complex conjugation. While this preserves the structure in classical mechanics or certain formal scenarios, it fails to maintain quantum mechanical consistency, leading to unbounded Hamiltonians and violating Wigner’s theorem \cite{Wigner}
 requiring symmetry representations to be unitary or anti-unitary.

In the unitary time reversal framework, the operator acts linearly without complex conjugation, and as a consequence, it reverses the signs of energy eigenvalues rather than interchanging initial and final states as in the anti-unitary picture. This reversal implies that under a unitary time reversal transformation, states with positive energy map to states with negative energy, effectively extending the energy spectrum below the usual positive energy bound. While negative energies are generally considered problematic in quantum mechanics—due to issues like vacuum instability—recent studies suggest that negative energy solutions can be accommodated if the concept of mass is simultaneously extended to include negative mass states. Such a scenario appears in modified relativistic quantum theories \cite{Petit1} that explore unitary time reversal as a fundamental symmetry.

The negative energy interpretation under unitary time reversal also alters the transformation properties of fundamental fields and operators. For instance, scalar field quanta exhibit energy reversal while momentum remains invariant, and spin eigenvalues may flip sign accordingly, preserving the overall consistency of quantum statistics. Moreover, the geometric and physical interpretation of velocity and mass become subtler, as the codependence on the time coordinate reversal no longer directly determines these dynamical quantities in the same way as the conventional anti-unitary case.

This conceptual framework offers potential resolutions to paradoxes traditionally associated with time reversal, such as the “running backward movie” analogy, by interpreting time reversal not as a literal reversal of event ordering but as a reversal of energy signs within a globally forward-flowing physical time. Consequently, unitary time reversal provides a mathematically consistent and physically meaningful alternative symmetry operation, especially in non-orientable spacetimes or exotic quantum gravitational scenarios where standard anti-unitarity may be questioned.

The unitary time reversal operator introduces negative energy states naturally into the quantum framework, challenging conventional assumptions and expanding the landscape of possible symmetries. This perspective enriches our understanding of discrete symmetries in quantum theory and opens new theoretical pathways in the study of quantum fields, spacetime topology, and fundamental time-reversal invariance.

\section{ Non-Orientable Black Hole Connecting Two Time-Reversed Spacetimes }

Wormholes in advanced gravitational frameworks, such as scalar-tensor gravity \cite{Lobo} and $PT$-symmetric theories \cite{Zejli}, exhibit exotic properties that challenge conventional notions of spacetime orientation and causality. In scalar-tensor gravity, wormholes can connect regions with reversed temporal flows, while $PT$-symmetric wormholes function as portals linking two spacetime sheets where physical time progresses in opposite directions. A particle traversing the throat of such a $PT$-symmetric wormhole undergoes a parity-time ($PT$) transformation, effectively experiencing a combined time reversal ($t \rightarrow -t$) and spatial inversion ($\vec{x} \rightarrow -\vec{x}$). This $PT$-symmetry implies a profound geometric consequence: the resulting global spacetime manifold becomes non-orientable, characterized by closed curves that return a traveler to the starting point but with reversed time and spatial orientations \cite{Hawking}.

This non-orientability aligns with fundamental features of non-orientable spacetimes, where certain paths—known as orientation-reversing loops—flip the handedness of space and the arrow of time upon closure. In the wormhole context, the throat acts as the critical junction enforcing this reversal, akin to gluing two mirror-image spacetime sheets along a timelike boundary.

To bolster this argument further, recent work on non-orientable BTZ black holes provides a precise analogy in lower-dimensional AdS gravity. In \cite{Racorean}, the eternal BTZ geometry is derived by interchanging space and time roles inside the horizon, yielding an interior metric dual to a thermofield double state. Combining interior and exterior thermofield doubles produces a full spacetime with Klein bottle-like non-orientable topology \cite{Tang, Hong}, where paths across the horizon link asymptotically AdS regions with opposite orientations, inducing $PT$ flips analogous to wormhole throats. 

These models collectively demonstrate how such structures enforce orientation-reversing identifications, with implications for time-reversed information flow, and exotic matter stabilization. The defining feature unifying these wormhole classes lies in the $PT$ symmetry transformation undergone by particles crossing the throat, explicitly combining time reversal (forward temporal evolution becomes backward) and spatial inversion (all spatial coordinates flip sign). This combined operation belongs to the improper component of the Lorentz group $O(1,3)$ rather than the proper subgroup $SO(1,3)$, producing a definitive change of orientation that renders the wormhole spacetime consistent with non-orientable topology. For a test particle entering from the left asymptotic sheet with future-directed proper time, emergence on the right sheet occurs with past-directed proper time relative to the throat frame. Momentum vectors reverse direction completely, spin polarization flips absolutely, and the particle's four-velocity transforms from future-pointing to past-pointing—all consequences of the throat's holonomy matrix explicitly implementing the $PT$ operation.

In scalar-tensor wormholes \cite{Lobo}, this manifests through discontinuity in the scalar field value across the throat; $PT$-symmetric configurations \cite{Zejli} achieve it through balanced complex metric sectors; BTZ non-orientable geometries \cite{Racorean} realize it via explicit metric signature interchange in the maximal extension of the spacetime geometry. The non-orientability becomes mathematically manifest: closed loops encircling the throat accumulate orientation-reversing holonomy, precisely analogous to the Mobius strip's single boundary surface where traversal returns the explorer to the origin facing the opposite direction.

We mention here also the non-orientable wormholes \cite{Dokuchaev} that exhibit only space non-orientability. 

The $PT$-induced orientation reversal produces the most striking quantum mechanical consequence: particles crossing from one side to the other unitarily change their energy from positive to negative.  In quantum mechanics, energy acts as the generator of time translations, appearing in the phase factor $e^{-iEt/\hslash}$ of the wave function (where $\hslash$ is the reduced Planck's constant). Reversing time—equivalent to replacing $t\to-t$—inverts this phase to $e^{iEt/\hslash}$, which mathematically maps positive energy states ($E>0$) to negative energy states ($E<0$). Since energy governs quantum phase evolution, time reversal flips the phase accumulation direction, converting forward-propagating positive energy wave packets into backward-propagating negative energy configurations.

Consider a spin particle incident on the throat with positive rest energy and upward spin projection. Upon emergence on the other side, it possesses definite negative energy with fully reversed spin direction. This unitary transformation maintains real energy eigenvalues in unbroken $PT$-symmetric phases despite the spectrum inversion. Thus, in summary, a spin particle that traverses such a wormhole undergoes a unitary $PT$ that corresponds to the transformations: 

\begin{equation}
\vec{x} \rightarrow -\vec{x}, \vec{p} \rightarrow -\vec{p}, E \rightarrow -E, m \rightarrow -m, i \rightarrow i
\end{equation}

which differs from the traditional anti-unitary $PT$:

\begin{equation}
\vec{x} \rightarrow -\vec{x}, \vec{p} \rightarrow \vec{p}, E \rightarrow E, i \rightarrow -i
\end{equation}

that, as we will see later, corresponds to orientable spacetimes.

This energy reversal process finds profound parallels with relativistic fermions described by the Dirac equation, where the spectrum naturally divides into positive and negative energy continua. An electron traversing such a wormhole throat effectively emerges as a positron analog: momentum completely reversed, helicity formally preserved but absolute spin direction flipped, energy sign inverted. This mirrors the Feynman-Stückelberg interpretation \cite{Feynman, Stueckelberg}
 where positrons propagate backward in time as negative energy electrons, with the throat functioning as an effective $CPT$ transformation interface that realizes charge conjugation through the energy sign change itself.

Spacetime orientability provides the conceptual framework clarifying these phenomena. An orientable spacetime admits globally consistent definitions of "handedness" through reduction of the frame bundle to the proper orthochronous Lorentz group, maintaining a uniform arrow of time essential for unambiguous particle-antiparticle classification and causal order preservation. Positive helicity fermions remain distinct from negative helicity states across the entire manifold, future-directed worldlines maintain global consistency, and chirality provides stable quantum number assignment.

Non-orientable spacetimes, by contrast, lack this global consistency, permitting closed loops where orientation reverses upon complete traversal—precisely analogous to the Mobius strip's topology in two dimensions, where a geodesic path returns the traveler to the origin atop the antipodal "side." In four-dimensional Lorentzian manifolds, non-orientability manifests independently in spatial dimensions (producing parity-flipping holonomy that confuses chiral states globally) or temporal dimensions ( producing a flip in the time direction), or through mixed combinations yielding full $PT$ holonomy. The Hawking-Ellis \cite{Hawking} classification systematically distinguishes these cases, each carrying profound physical implications for quantum field theory construction.

\section{ Unitary time reversal operator on non-orientable spacetimes and anti-unitary on orientable spacetimes }

We have seen that implementation of time reversal symmetry in quantum mechanics is closely tied to the global topological property of spacetime known as orientability. Traditionally, in quantum theory formulated on orientable spacetimes, time reversal is represented by an anti-unitary operator. This anti-unitary nature arises because time reversal not only reverses the direction of time but also requires complex conjugation to maintain the fundamental algebraic and physical properties of quantum states and observables. The anti-unitarity is fundamental to the Wigner theorem's assertion that symmetry operations on quantum states must be either unitary or anti-unitary, with time reversal falling into the latter category due to its antilinear properties.

Contrastingly, in non-orientable spacetimes—where a global, consistent time orientation does not exist—the nature of the time reversal operator can fundamentally differ. Non-orientability implies the presence of loops within spacetime that reverse temporal orientation upon traversal, akin to how a Mobius strip reverses spatial orientation globally. In such a topology, the reversal of time's arrow is encoded in the geometry itself rather than enforced algebraically by complex conjugation on the quantum states. This opens the possibility for time reversal symmetry to be realized by a purely unitary operator that is linear, without the need for the anti-linear complex conjugation typically required in orientable contexts.

The unitary representation of time reversal in non-orientable spacetimes reflects a topologically induced inversion of temporal orientation, rendering the complex conjugation superfluous. In effect, the non-orientable structure performs the role traditionally played by the anti-unitary operator's antilinearity. As a consequence, the quantum mechanical time-reversed partner state arises purely through a unitary transformation consistent with the altered global topology.

The distinction between unitary and anti-unitary time reversal has significant implications. In orientable spacetimes, anti-unitarity guarantees that the reversal of time flips momenta, spin, and complex phases consistently and preserves the probability interpretation of quantum mechanics. In non-orientable spacetimes, the topological reversal of time orientation enables a unitary time reversal operator that acts without conjugation yet still achieves the physical inversion of temporal direction via the spacetime’s global properties. Such generalized time reversal symmetry potentially leads to novel quantum effects, especially in gravitational systems with nontrivial topology where non-orientability naturally arises.

\section{ Time Reversal in the Dirac and Schrödinger Equations: Unitary versus Anti-Unitary Realizations}

The interplay between the mathematical form of quantum wave equations and the topological properties of the underlying spacetime is essential in understanding discrete symmetries such as time reversal. A striking manifestation of this connection is found in the different realizations of the time reversal operator in the Dirac and Schrödinger equations—realisations that align closely with the orientability or non-orientability of the spacetime manifold.

The Dirac equation, fundamentally relativistic and intrinsically incorporating spin-½ degrees of freedom, admits a consistent unitary time reversal operator in the context of non-orientable spacetimes. The non-orientability, which corresponds to the absence of a global, consistent time direction, naturally encodes time reversal symmetries into its topological structure. In such spacetimes, the unitary time reversal operator acts in a manner that reverses the direction of time without requiring complex conjugation. This operator exchanges positive and negative energy solutions, reflecting the nontrivial topology's encoding of the reversal of time's arrow. The algebraic framework of the Dirac equation—with its gamma matrices and spinor representation—supports these unitary transformations, allowing the equation to remain invariant under this symmetry. 

Let us consider a spin particle that traverses form one side of the $PT$-wormhole to the other. The positive energy solutions of the particle

\begin{equation}
\Psi^+_1=
\sqrt{\frac{E+m}{2E}}
e^{i(px-Et)}
\begin{pmatrix}
1 \\
0  \\
\frac{p_3}{ E+m } \\
\frac{p_1+ip_2}{ E+m}
\end{pmatrix}
,
\Psi^+_2=
\sqrt{\frac{E+m}{2E}}
e^{i(px-Et)}
\begin{pmatrix}
0 \\
1 \\
\frac{p_1-ip_2}{E+m} \\
-\frac{p_3}{E+m}
\end{pmatrix}
\end{equation}

changes under the transformations in Eq. (8) as,

\begin{equation}
\Psi^-_1=
\sqrt{\frac{E+m}{2E}}
e^{-i(px-Et)}
\begin{pmatrix}
\frac{p_3}{ E+m } \\
\frac{p_1+ip_2}{ E+m}  \\
1 \\
0
\end{pmatrix}
,
\Psi^-_2=
\sqrt{\frac{E+m}{2E}}
e^{-i(px-Et)}
\begin{pmatrix}
\frac{p_1-ip_2}{E+m} \\
-\frac{p_3}{E+m} \\
1 \\
0
\end{pmatrix}
\end{equation}

Thus, the anti-fermions in Eq, (11) have negative energies and masses.

On the contrary, the Schrödinger equation, a non-relativistic equation describing quantum dynamics on orientable spacetimes, demands an anti-unitary time reversal operator. The Schrödinger dynamics depend explicitly on the imaginary unit $i$ within the time evolution operator $e^{-iHt/\hbar}$. Reversing time, i.e., $t \to -t$, requires the conjugation operation to invert the sign of $i$, which cannot be achieved by any purely linear unitary operator. Hence, complex conjugation is indispensable, and the time reversal operator must be anti-unitary in orientable spacetimes, where a global temporal orientation is well defined and preserved throughout the manifold. This anti-unitary structure ensures correct physical behavior such as momentum inversion, preservation of canonical commutation relations, and probability conservation. 

This dichotomy reveals a fundamental link between spacetime topology and quantum symmetry operations: non-orientable spacetimes, characterized by lack of time orientability
, accommodate a unitary time reversal operator in relativistic quantum theories like the Dirac equation, while orientable spacetimes enforce the traditional anti-unitary time reversal in non-relativistic theories exemplified by the Schrödinger equation. This relationship underscores that the algebraic form of quantum dynamics and the geometric-topological structure of spacetime are intertwined in determining how fundamental symmetries manifest.

Furthermore, the allowance of negative energy states in the Dirac framework under unitary time reversal dovetails with the non-orientable topology’s inherent time inversion. These negative energy states, problematic in standard interpretations, acquire natural meaning as partners under the topological time reversal symmetry, potentially opening new theoretical avenues in quantum field theory on exotic spacetimes and quantum gravity.

\section{Conclusion}

Time reversal symmetry occupies a uniquely subtle role in quantum mechanics, differing fundamentally from other symmetries such as spatial parity or rotations. Unlike these, which are faithfully represented by unitary operators acting linearly on a Hilbert space, time reversal requires an anti-unitary operator—one that combines a unitary transformation with complex conjugation. This anti-unitarity is indispensable because it ensures the correct inversion of the imaginary unit $i$ present in canonical commutation relations and dynamical evolution. Without this complex conjugation, time reversal implemented as a purely unitary operation cannot consistently flip momentum while preserving the algebraic and probabilistic structure of quantum mechanics.

The anti-unitary time reversal operator guarantees the physical consistency of quantum theory by preserving essential structures such as the canonical commutation relation $[x,p]=i\hslash$, ensuring momentum reverses sign while position remains invariant. It also prevents unphysical consequences like negative energy spectra from violating stability, since the energy operator behaves correctly under anti-unitary transformation. Wigner’s theorem firmly establishes that all fundamental symmetries in quantum mechanics must be represented by either unitary or anti-unitary operators, with time reversal belonging to the latter class due to its unique properties.

Despite this necessity, recent research motivated by quantum gravity and black hole physics uncovers scenarios where the conventional anti-unitary picture can be generalized. In particular, spacetimes with non-orientable topology—a feature characterized by the impossibility of defining a consistent global temporal orientation—permit an alternative realization of time reversal symmetry using a purely unitary operator. This possibility emerges naturally in particular wormhole classes that connect regions with reversed temporal flows such as wormhole in scalar–tensor gravity \cite{Lobo}, PT-symmetric wormholes \cite{Zejli} or non-orientable BTZ black holes \cite{Racorean}. Here, time reversal emerges as a topological operation encoded in the manifold itself, bypassing the need for complex conjugation traditionally required in orientable spacetimes.

This remarkable insight demonstrates that the global topological characteristics of spacetime fundamentally influence the algebraic structure of quantum symmetries. Orientable spacetimes, characterized by a fixed global arrow of time, necessitate an anti-unitary time reversal operator to maintain the consistency of quantum mechanics, a condition naturally realized in the Schrödinger framework. Conversely, non-orientable spacetimes accommodate unitary time reversal operators consistent with negative energy states and a topological inversion of time’s arrow, as manifested in relativistic systems described by the Dirac equation.

The interplay between time reversal symmetry and spacetime topology not only refines our understanding of discrete quantum symmetries, but also raises profound implications for quantum gravity, black hole thermodynamics, and the nature of time itself. The presence of negative energy states and reversed temporal orientation within non-orientable spacetimes offers novel pathways to re-examine causality, information paradoxes, and the emergence of time’s arrow from a topological perspective.

\end{document}